\documentclass[mathleft]{an} 
\usepackage{graphicx}
\usepackage{times}
\begin{document}

\Pagespan{1}{}
\Yearpublication{2010}%
\Yearsubmission{2010}%
\Month{}%
\Volume{}%
\Issue{}%

\title{Inhomogeneous Power Distribution in Magnetic Oscillations}

\author{Kiran Jain\thanks{Corresponding author:
  \email{kjain@noao.edu}\newline}
\and S. C. Tripathy \and S. Kholikov \and F. Hill
}
\titlerunning{Power in Magnetic Oscillations}
\authorrunning{K. Jain et al.}
\institute{GONG Program, National Solar Observatory, 
950 N Cherry Avenue, Tucson, AZ 85719 USA
}
\received{ }
\accepted{ }
\publonline{later}


\abstract{%
   We apply ring-diagram analysis and spherical harmonic decomposition methods to compute
 3-dimensional power spectra of magnetograms obtained by the Global Oscillation Network
 Group (GONG) during quiet periods of solar activity. This allows us to
 investigate the power distribution in acoustic waves propagating in localized directions on
 the solar disk.  We find evidence of the presence of five-minute oscillations in
 magnetic signals that suggests a non-homogeneous distribution of
 acoustic power. In this paper, we present our results on the asymmetry
 in oscillatory power and its behaviour as a function of frequency, time
 and magnetic field strength. These characteristics are compared with simultaneous
 velocity measurements.}

\maketitle

\section{Introduction}

The nature of the propagation of acoustic waves in the solar atmosphere is important 
for understanding the interaction of these waves with magnetic fields that modify 
the surface amplitude of the propagating waves. The absorption of acoustic waves by 
sunspots  has been discussed using various data sets and techniques (e.g. Braun, 
La Bonte and Duvall 1987, Lites et al. 1998, Norton et al. 1999, Ulrich 1996). The 
studies using the techniques of local-helioseismology have also shown a significant 
decrease in oscillatory power in  localized regions of strong magnetic field (Chou 
et al. 2009, Howe et al. 2004, Rajaguru, Basu and Antia 2001).  The observed modulation 
in power is also affected by neighboring regions and this ``neighborhood effect'' is  
discussed in detail by Nicholas, Thompson and Rajaguru (2004).  Observations further 
suggest that the amplitude is suppressed at low-frequencies while it is enhanced 
around magnetic field regions above 5 mHz (Jain \& Haber 2002). 

In this paper, we present characteristics of the oscillations observed in  magnetic 
and velocity signals during quiet solar activity using simultaneous full-disk measurements 
from Global Oscillations Network Group (GONG). The origin of five-minute oscillations 
in magnetograms is believed to be cross talk between Doppler velocity and Zeeman 
splitting. Since the presence of strong magnetic fields modulates the power, quiet 
periods provide clues to the me-chanism of how the waves propagate in the absence 
of the field. For a comprehensive comparison of results, two different techniques, 
ring-diagram analysis and spherical harmonic decomposition, are applied.

\section{Data and Analysis}

We use high-cadence (60 s) line-of-sight continuous magnetograms from GONG. These 
magnetograms were obtained in the Ni 6768 $\AA$ spectral line with a spatial pixel 
size of approximately 2.5 arc-sec, and have a noise level of 3~G per pixel. To study 
the acoustic power distribution in the quiet Sun,  we choose a period of minimal 
activity in both front- and far-sides. Hence we analyze data during 2008 August, 
which recorded more than 94\%  spotless days.  We also use simultaneous Dopplergrams 
from the same network to compare the characteristic of oscillatory power in two 
observables. We treat these images locally by  applying ring-diagram technique 
and globally by using the spherical harmonic decomposition method to calculate 
power spectra. 

For the ring-diagram method, we used a grid of 128$\times$  128 pixels with
 spatial resolution of 0$^{\mathrm{o}}$.25 at disk center. The regions of
 32$^{\mathrm{o}}$ square (apodized to 30$^{\mathrm{o}}$ diameter) were remapped
 and tracked for 1440 min
 using the surface rotation rate of Snodgrass. The  Fast Fourier Transformation
 (FFT) was then applied to the tracked cube to calculate 3-dimensional power spectrum 
($k_x$, $k_y$, $\omega$). In this method, 
 the oscillation power within the spectrum is distributed along curved surfaces that,
 when cut at constant frequency, appear as a set of nested rings, each corresponding to
 a mode of different radial order $n$. Figure~1 shows rings obtained at 3.333 mHz for
 both magnetic and velocity signals. 

\begin{figure}
\begin{center}
\includegraphics[width=40mm]{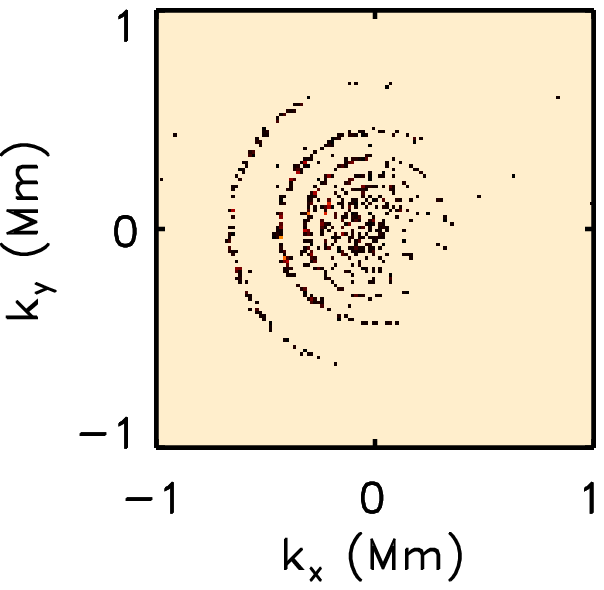}
\includegraphics[width=40mm]{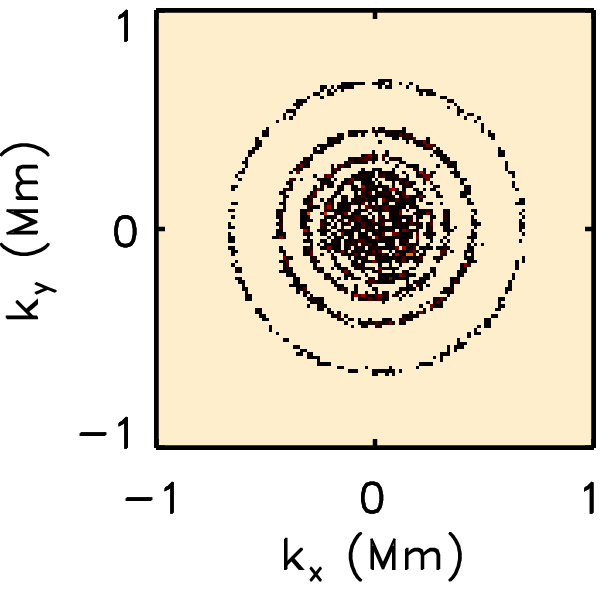}  
\caption{Cross sectional cuts of a three-dimensional ring-diagram power spectra at a 
temporal frequency 3.333 mHz using {\it (Left)} magnetograms, and {\it(Right)} 
Dopplergrams. }
\label{label1}
\end{center}
\end{figure}

\begin{figure}
\begin{center}
\includegraphics[width=80mm]{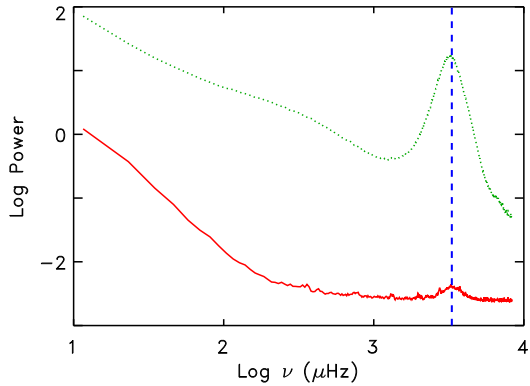}  
\caption{Azimuthally-averaged 2D power spectra as a function of temporal frequency $\nu$
for magnetic (solid/red) and velocity (dotted/green) oscillations. The vertical 
dashed line corresponds to 3.3 mHz (5-min oscillation).}
\label{label2}
\end{center}
\end{figure}

For the spherical harmonic decomposition method, regions of 120$^{\mathrm{o}}$ in 
diameter centered at the disk center were selected and  remapped into sin(lat)-long 
co-ordinates. The spherical harmonic decomposition was applied to produce a time 
series of coefficients.  The obtained time series were filtered with a Gaussian 
filter of FWHM = 2.5 mHz centered at 3.3 mHz and the FFT was applied to produce
 power spectrum. 

\section{Results}
\subsection{Ring-diagram analysis}
Figure~1 shows the cuts of 3D spectra in ($k_x$, $k_y$)  plane for magnetic 
and velocity oscillations  at a constant frequency ($\nu$ = 3.333 mHz) for 
2008 August 18.  It is clearly seen that the power around the rings in magnetic 
oscillations is different from the velocity oscillations.  These slices provide 
information about the characteristics of the propagated waves. The approximate 
relationship between $k_x$, $k_y$, $\ell$ and $m$ are $k_x \approx$ $\sqrt{m^2}$/$R$ 
 and $k_y \approx$ $\sqrt {\ell^2 -m^2}$/$R$, where $R$ is the solar radius
 and other symbols have their usual meanings. The total wave number, 
$k$, is $\sqrt {\ell(\ell+1)}$/$R$. 

 The presence of partial rings in magnetic oscillations suggests anisotropic
 distribution of acoustic power in propagating waves. These were first reported by 
Hill et al. (2008) where quiet, sunspot and network regions were analyzed.  In
 this analysis, it was conjectured that the power in quiet region might have been 
suppressed by the presence of neighboring sunspot. Since our analysis is confined to
 quiet days and there is no visible solar activity for several days, we do not expect
 the observed suppression to be arising  from the ``neighborhood effect''.
We also notice that the power in magnetic oscillations
 is relatively weak. This difference is clearly visible in Figure~2 where we show
 azimuthally averaged power spectra for both the observables. We do see a bump in magnetic
 oscillations at 3.33 mHz but it is less prominent compared to the velocity
 oscillations due to low signal-to-noise ratio.  A similar trend is observed 
for all days considered in this analysis. As an example, we show in Figure~3 
the variation of normalized power in ring-diagram analysis with azimuth for 
three consecutive days. In an earlier analysis using data from NaI D line 
by Magneto-Optical Filter (MOF) instrument, Moretti et al. (2003) also found a 
significantly low signal-to-noise ratio in magnetic oscillations. We also 
find a weak anisotropy in power distribution in velocity rings.

\begin{figure}
\begin{center}
\includegraphics[width=80mm]{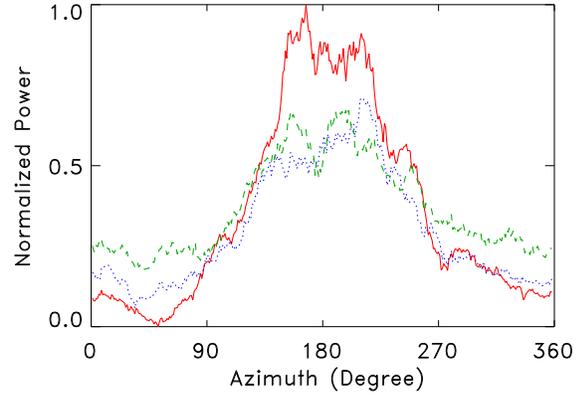}
\caption{Acoustic power distribution in magnetic oscillations as a function 
of azimuth at a constant frequency 3.3 mHz for three sequential days of 
observations in 2008; August 16 (dashed/green), August 17 (dotted/blue) and 
August 18 (solid/red). }
\label{label3}
\end{center}
\end{figure}

\subsection{Spherical harmonic decomposition method}
Here we split the total power spectrum into two components corresponding to 
positive- and negative-$m$ coefficients. In Figure~4, we show the variation 
of these components with frequency for magnetic oscillations. In all cases, 
we find that the waves propagating in the direction of solar rotation 
(represented by $+m$ coefficients) have more power than the retrograde waves. 
A similar plot for velocity oscillations is shown in Figure~5, but we obtain 
comparable power in both directions. Hence, in this method again, we find 
the acoustic power in waves seen in magnetic field propagating in westward 
direction is higher than the eastward direction. This indicates that a 
portion of power is lost when waves propagate against the direction of 
solar rotation, however this effect is prominent in magnetic oscillations
 due to the low signal-to-noise ratio. These results support the findings 
obtained with ring-diagram technique as discussed earlier.

\begin{figure}
\begin{center}
\includegraphics[width=80mm]{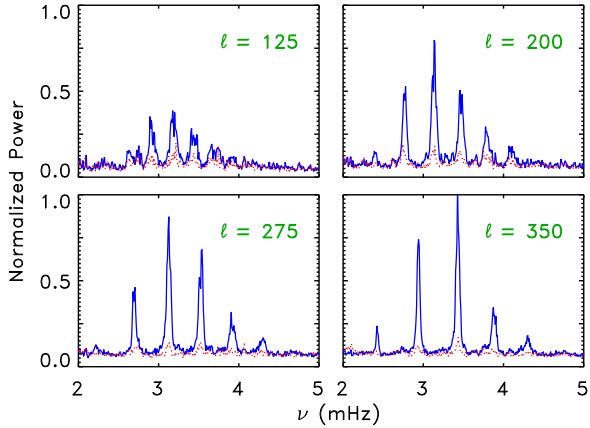}  
\caption{Power spectra obtained from spherical harmonic decomposition method 
at different values of $\ell$ for magnetic oscillations. The solid (blue) and 
dotted (red) lines are for power in waves propagating in 
westward and eastward directions respectively. Results are shown for 2008 
August 18.}
\label{label4}
\end{center}
\end{figure}

\begin{figure}
\begin{center}
\includegraphics[width=80mm]{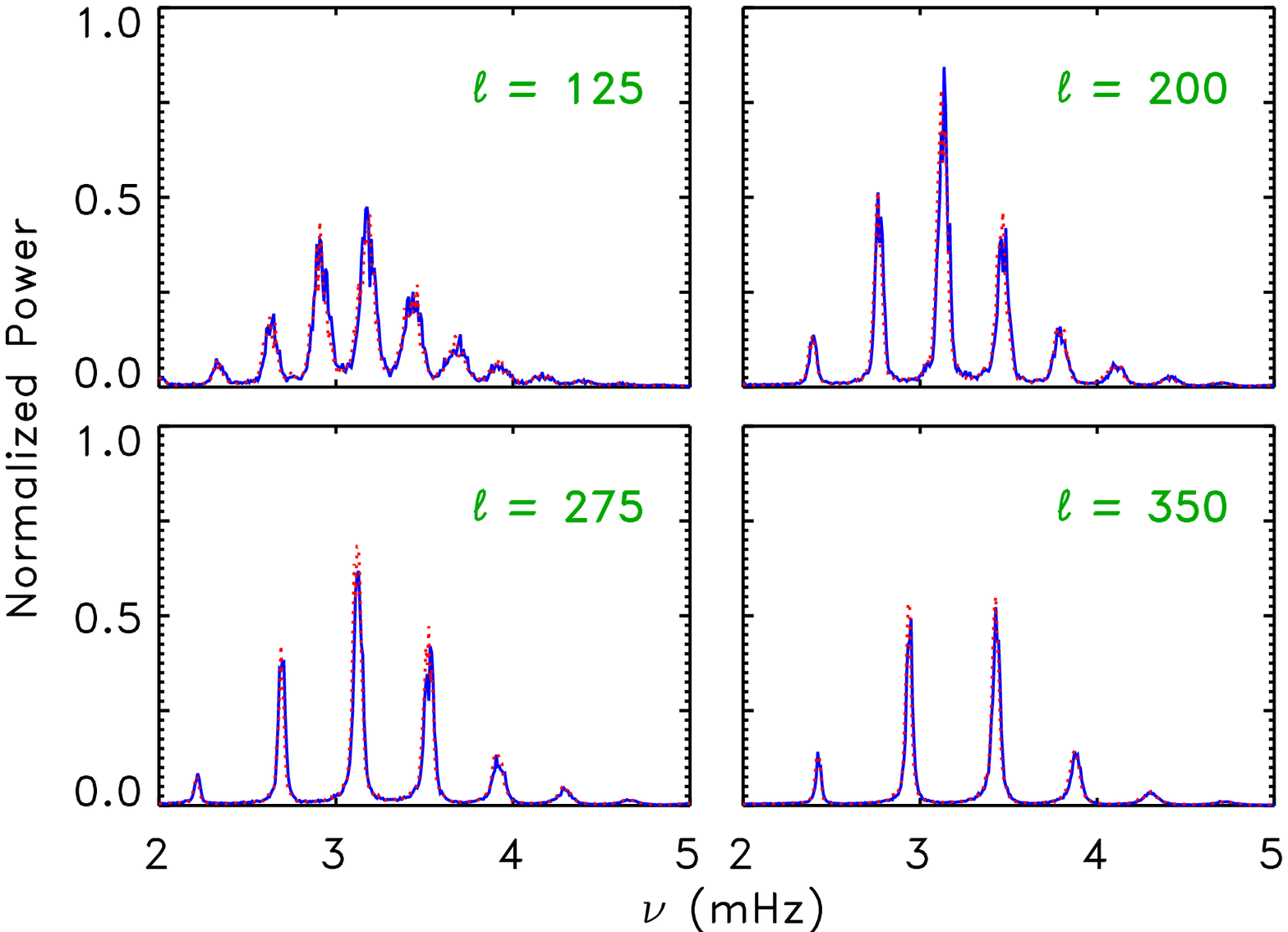}  
\caption{Same as Figure 4 but for velocity oscillations.}
\label{label5}
\end{center}
\end{figure}
\subsection{Temporal variability in anisotropy}

\begin{figure}[t]
\begin{center}
\hskip 0.5cm
\includegraphics[width=75mm]{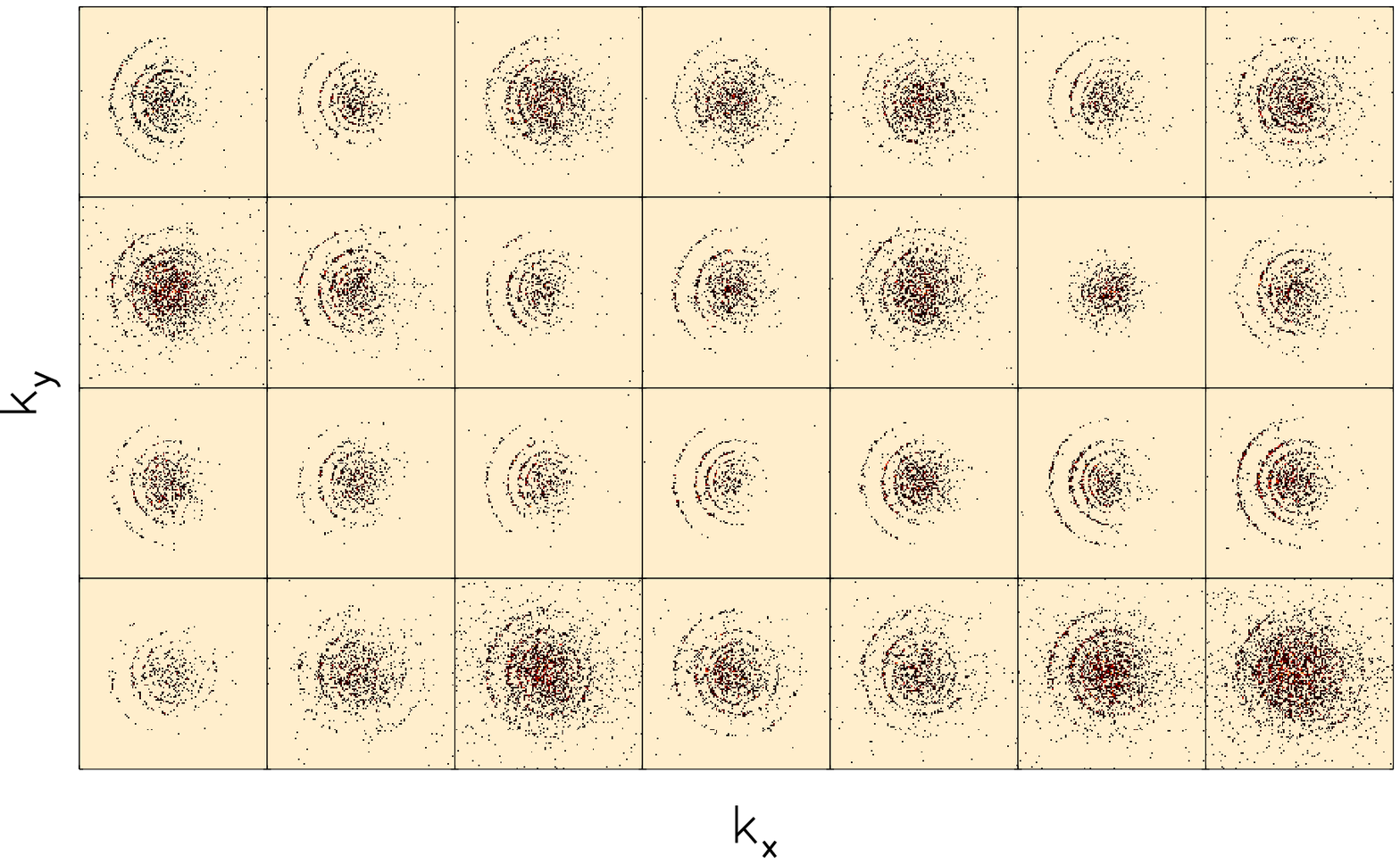}
\vskip -0.25in
\includegraphics[width=80mm]{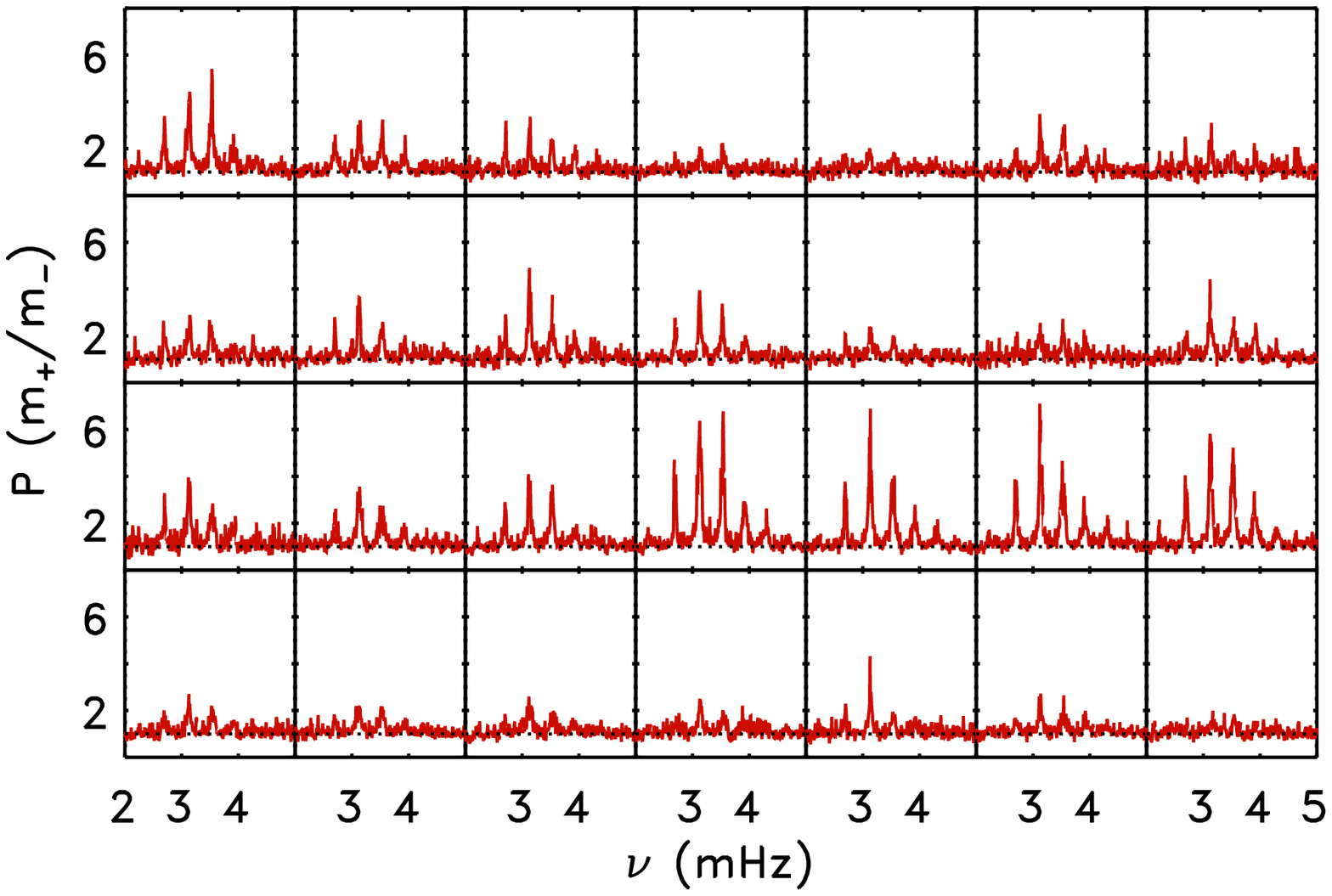}
\caption{ Magnetic oscillations at disk center for the period 2008 
August 1-28: ({\it Top}) Cross-sectional cuts of three-dimensional 
ring-diagram power spectra at 3.333 mHz. The scales on x- and y-axes 
are same as in Figure~1.
({\it Bottom}) The
asymmetry parameter as a function of frequency obtained from spherical
 harmonic decomposition method for $\ell$ = 275. }
\label{label6}
\end{center}
\end{figure}
\begin{figure}
\begin{center}
\hskip 0.5cm
\includegraphics[width=75mm]{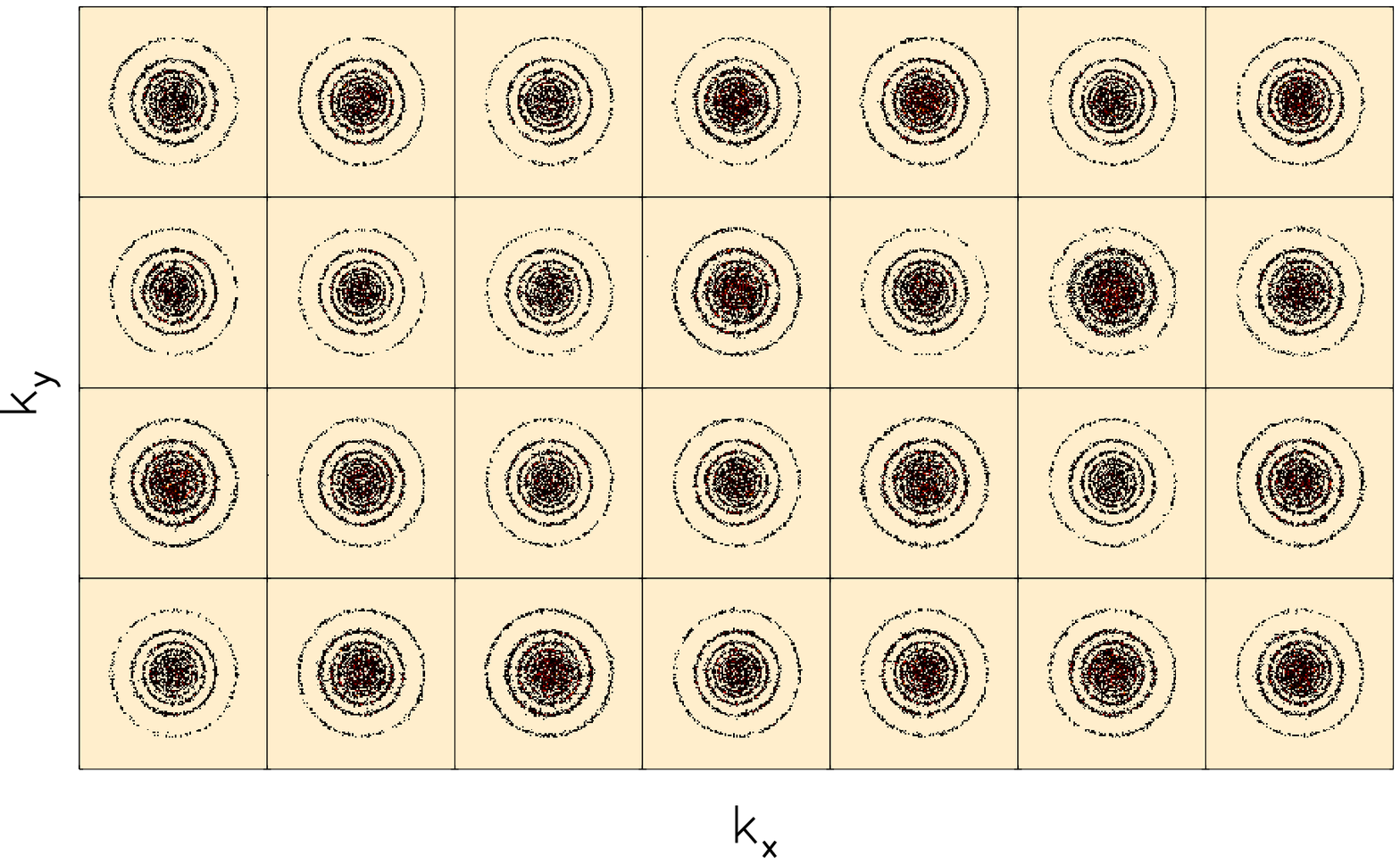}
\vskip -0.25in
\includegraphics[width=80mm]{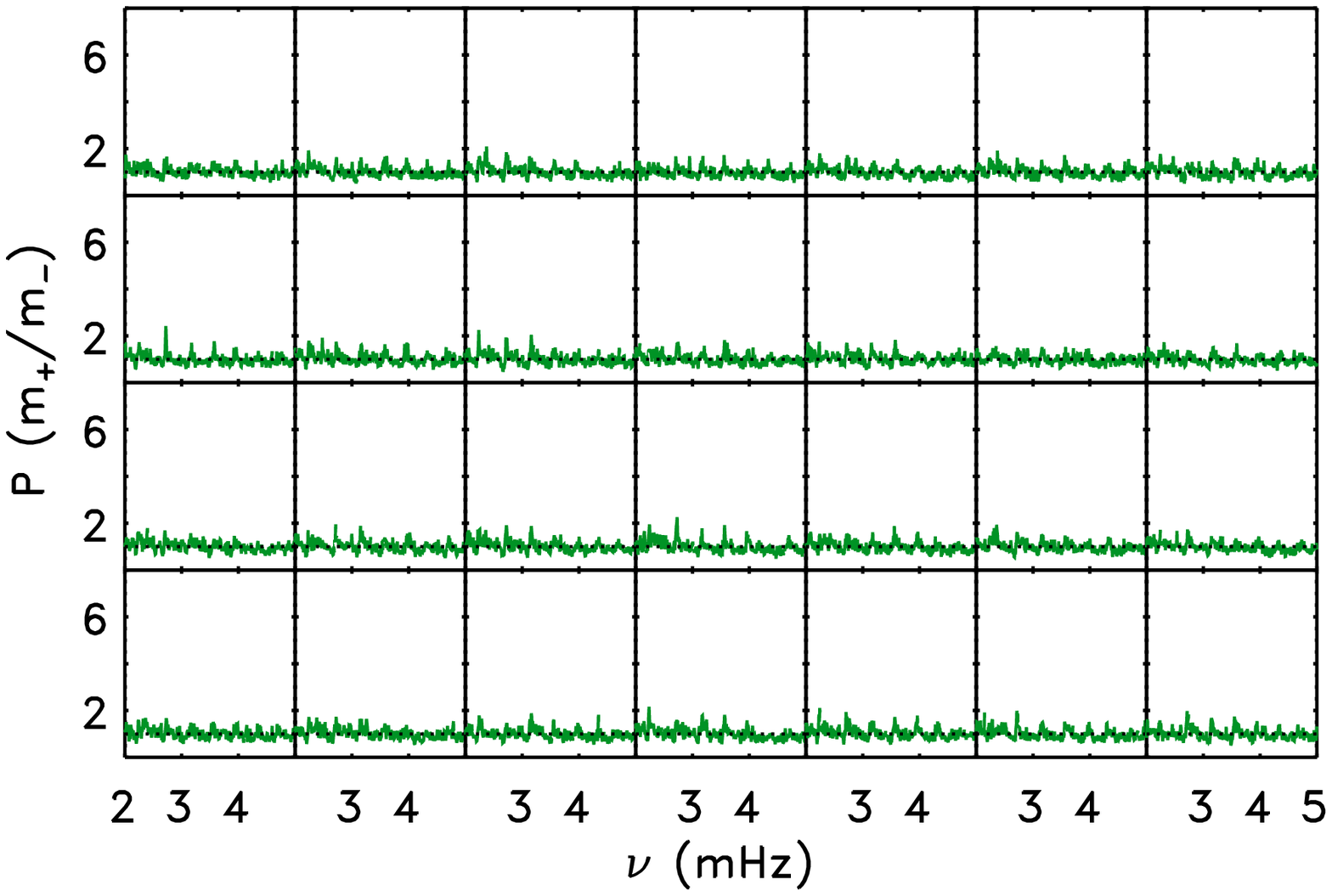}
\caption{Same as Figure 6 but for velocity oscillations.}
\label{label7}
\end{center}
\end{figure}

To study the temporal variability in anisotropy, we plot in Figure~6 the 
3-dimensional power spectra at $\nu$ = 3.333 mHz for the period 2008 August 1-28. Although 
the anisotropy is present in all cases, there are variations from day to day. 
It is well known that the duty cycle of observations plays an important role 
in helioseismic techniques and affects the observations. Clearly, a low duty 
cycle will produce noisy rings. A detailed analysis of the effect of duty cycle 
on anisotropy is in progress and will be published elsewhere.

 To quantify the anisotropy we define an asymmetry parameter,  $P (m_+/m_-)$, as the
 ratio of the power in waves propagating in westward (prograde) and
 eastward (retrograde) directions.  These values obtained from the spherical
 decomposition method are plotted in Figure~6. It is evident 
that the asymmetry parameter varies with frequency and has significantly 
higher values for the peaks in power spectra. These are also found to 
vary with time. We notice a close correspondence between asymmetries obtained 
with both the methods. Figure~7 shows the power distribution in the velocity 
oscillations. As discussed above, there is a weak inhomogeneity in
 velocity oscillations but it is not as significant as in magnetic oscillations.
 Hence our analysis of power spectra for waves in two different directions clearly
 indicates that the propagation is affected by the solar rotation that attenuates the
 amplitude when waves propagates against this direction.

To study the effect of the magnetic field strength on the asymmetry 
parameter,  Figure~8 shows $P (m_+/m_-)$ as a function of magnetic index 
for four different frequency bands. We find a systematic variation in
asymmetry parameter with frequencies and maximum values are achieved 
for the 3.5 mHz band. However, we do not find any significant correlation 
between the asymmetry parameter and magnetic field strength. 
A statistically meaningful analysis for this variation is in 
progress and we believe that such studies will provide 
important clues on the absorption of power in magnetized regions.

\begin{figure}
\begin{center}
\includegraphics[width=80mm]{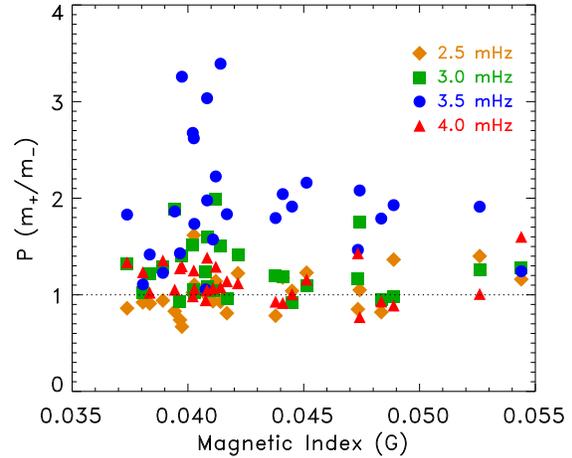}
\caption{The variation of asymmetry parameter for $\ell$=275  with 
magnetic field index for four different frequency bands. }
\label{label8}
\end{center}
\end{figure}

\section{Summary}

Using simultaneous Dopplergrams and magnetograms obtained from GONG, 
we find evidence of the presence of 5-minute oscillations in magnetic 
signals.  The oscillations in these magnetograms are believed to arise 
due to cross talk between Doppler velocity and Zeeman splitting. The 
important findings in this work are summarized below;

\begin{enumerate}
\item The signal-to-noise ratio in magnetic rings is low compared to 
the velocity rings.
\item The anisotropy in magnetic rings appear in two quadrants 
that correspond to waves propagating in the retrograde direction. 
Thus, in the quiet sun the effect is mainly due to solar rotation
which shifts the frequency of one half of the ring, thereby 
decreasing its amplitude.
\item The asymmetry parameter varies with frequency and the maximum 
asymmetry is obtained in 5-min oscillation band. 
 \item Our analysis do not show any significant correlation 
between asymmetry parameter and magnetic field.
\item We do find anisotropy in the velocity rings, but it is much 
less prominent due to high signal-to-noise ratio. 
\end{enumerate}

\acknowledgements
 We wish to thank Irene Gonz{\'a}lez Hern{\'a}ndez for many useful 
discussions. This work was supported by NASA-GI grant NNG 08EI54I.
It utilizes data obtained by the Global Oscillation Network
Group (GONG) project, managed by the National Solar Observatory, which
is operated by AURA, Inc. under a cooperative agreement with the
National Science Foundation. The data were acquired by instruments
operated by the Big Bear Solar Observatory, High Altitude Observatory,
Learmonth Solar Observatory, Udaipur Solar Observatory, Instituto de
Astrof\'{\i}sico de Canarias, and Cerro Tololo Interamerican
Observatory.

\end{document}